\documentclass[12pt,onecolumn,final]{IEEEtran}
\newlength \figwidth
\usepackage{cite}
\ifCLASSINFOpdf
  \usepackage{graphicx,epstopdf}
  \graphicspath{{../Figures/}}
  \DeclareGraphicsExtensions{.eps}
\else
  \usepackage[dvips]{graphicx}
\fi
\usepackage[cmex10]{amsmath}
\usepackage[tight,footnotesize]{subfigure}
\usepackage{amssymb}
\usepackage{amsthm}
\usepackage{url}
\usepackage{dsfont}
\usepackage{tabulary}
\usepackage{multirow}
\usepackage{color}

\usepackage{color}
\usepackage{times}
\usepackage{verbatim}
\usepackage{floatflt}
\usepackage{enumerate}
\usepackage{array}
\usepackage{multicol,afterpage,wrapfig} 
\usepackage{tikz}
\usetikzlibrary{arrows,calc}
\allowdisplaybreaks

\begin{document}
\newcommand{\indic}{\mathds{1}}
\newcommand{\Herm}{\dagger}
\newcommand{\EPhi}{\mathbb{E}_{\Phi_e}}
\newcommand{\gMo}{\gamma_{M,o}}
\newcommand{\go}{\gamma_o}

\makeatletter
\def\thm@space@setup{\thm@preskip=0pt
\thm@postskip=0pt}
\makeatother

\newtheorem{Theorem}{\bf Theorem}
\newtheorem{Corollary}{\bf Corollary}
\newtheorem{Remark}{\bf Remark}
\newtheorem{Lemma}{\bf Lemma}
\newtheorem{Proposition}{\bf Proposition}
\newtheorem{Assumption}{\bf Assumption}
\newtheorem{Definition}{\bf Definition}
\title{Physical Layer Security in Downlink\\Multi-Antenna Cellular Networks}
\author{\normalsize Giovanni~Geraci, Harpreet~S.~Dhillon, Jeffrey~G.~Andrews, Jinhong~Yuan, and Iain~B.~Collings
\thanks{G.~Geraci and J.~Yuan are with the School of Electrical Engineering and Telecommunications, The University of New South Wales, Sydney, Australia (e-mail: g.geraci@student.unsw.edu.au, j.yuan@unsw.edu.au). H.~S.~Dhillon and J.~G.~Andrews are with the Wireless Networking and Communications Group (WNCG), The University of Texas at Austin,
TX (email: dhillon@utexas.edu and jandrews@ece.utexas.edu). I.~B.~Collings is with the Wireless and Networking Technologies Laboratory, CSIRO ICT Centre, Sydney, Australia (email: iain.collings@csiro.au). This work was done while the first author was with the WNCG, The University of Texas at Austin, TX. Manuscript updated: July 27, 2013.}
}
\maketitle
\vspace*{-0.8cm}
\thispagestyle{empty}
\begin{abstract}
In this paper, we study physical layer security for the downlink of cellular networks, where the confidential messages transmitted to each mobile user can be eavesdropped by both (i) the other users in the same cell and (ii) the users in the other cells. The locations of base stations and mobile users are modeled as two independent two-dimensional Poisson point processes. Using the proposed model, we analyze the secrecy rates achievable by regularized channel inversion (RCI) precoding by performing a large-system analysis that combines tools from stochastic geometry and random matrix theory. We obtain approximations for the probability of secrecy outage and the mean secrecy rate, and characterize regimes where RCI precoding achieves a non-zero secrecy rate. We find that unlike isolated cells, the secrecy rate in a cellular network does not grow monotonically with the transmit power, and the network tends to be in secrecy outage if the transmit power grows unbounded. Furthermore, we show that there is an optimal value for the base station deployment density that maximizes the secrecy rate, and this value is a decreasing function of the signal-to-noise ratio.
\end{abstract}

\IEEEpeerreviewmaketitle
\newpage
\section{Introduction}

Security is regarded as a critical concern in wireless multiuser networks. Due to its broadcast nature, wireless multiuser communication is very susceptible to eavesdropping, and it is essential to protect the transmitted information. The emergence of large-scale and dynamic networks imposes new challenges on classical
security measures such as network layer cryptography. To this end, exploiting the physical layer has been proposed as an alternative to achieve perfect secrecy without requiring key distribution and complex encryption/decryption algorithms \cite{Wyner75}. In the past few years, physical layer security has become a very active area of research, and has witnessed significant growth \cite{MukherjeeSurvey,Shiu11,LiangBook,LiuBook}.

\subsection{Motivation and Related Work}

Physical layer security for multi-user communications was first investigated by introducing the broadcast channel with confidential messages (BCC) \cite{Csiszar78}. In the multiple-input multiple-output (MIMO) BCC, a central base station (BS) with $N$ antennas simultaneously communicates to $K$ users which can act maliciously as eavesdroppers. The secrecy capacity region of a two-user MIMO BCC was studied, among others, in \cite{Ekrem12,Liu13,FakoorianJSAC}, under the assumption of perfect channel state information (CSI) available at the BS. The secrecy degrees of freedom, capturing the behavior of the secrecy capacity in the high signal-to-noise ratio (SNR) regime, were studied in \cite{Yang13,KobayashiSPAWC13} for the case when only delayed CSI is available at the BS. For larger BCC with an arbitrary number of malicious users, linear precoding based on regularized channel inversion (RCI) was proposed as a practical, low-complexity transmission scheme \cite{Geraci12}. In \cite{GeraciJSAC,GeraciLetter}, the authors employed random matrix theory tools to study the secrecy rate achievable by RCI precoding in the BCC under imperfect CSI and spatially correlated channels. In all these contributions, eavesdropping activity was assumed from the malicious users only. In practice, external nodes might be eavesdropping too.

The presence of external eavesdropping nodes and its effect on the secure connectivity in random wireless networks were studied in \cite{Haenggi08isit,Pinto12I}, where the authors investigated the secrecy communication graphs by employing stochastic geometry tools. It was shown in \cite{ZhouGanti11} that an improvement in the secure connectivity can be achieved by introducing directional antenna elements, whereas \cite{Zhou11TWC} investigated the throughput cost of security. The secrecy rates achievable in large ad hoc networks in the presence of colluding eavesdroppers and the scaling laws for secrecy capacity were derived in \cite{Pinto12II} and \cite{Koyluoglu}, respectively. The broadcast channel with confidential messages and external eavesdroppers (BCCE) was then introduced in \cite{GeraciTWC} to model a more general setting where both malicious users and randomly located external nodes can act as eavesdroppers. 

With a key exception of \cite{Wang13}, almost all the prior art focused on either an isolated cell or an ad hoc network, as discussed above. An attempt to study secrecy rate in the downlink of a cellular network is made in \cite{Wang13} by using tools from stochastic geometry. The current paper differs from and generalizes \cite{Wang13} in three key aspects: (i) while \cite{Wang13} considers single antenna transmission with orthogonal resource allocation, we consider a significantly generalized physical layer model with multiple transmit antennas serving multiple users with RCI based linear precoding, which may result in intra-cell interference, (ii) while \cite{Wang13} assumes that the interfering BSs are far away and that the inter-cell interference can be incorporated in the constant noise power, our model accounts for the exact inter-cell interference at the typical user, and (iii) while \cite{Wang13} assumes that only certain nodes in the network can eavesdrop without cooperation, we assume that all the users except the typical user, for which the secrecy rate and outage is computed, can cooperate to eavesdrop the transmitted messages meant for the typical user. Ignoring the three aspects above in the design of a physical layer security system, would make confidential communications vulnerable to secrecy outage caused by co-channel interference and inter-cell information leakage. For these reasons, it is of critical importance to extend the study of physical layer security to cellular networks, by taking into account the interference and the information leakage at cooperating malicious users.

\subsection{Approach and Contributions}

The main goal of this paper is to study physical layer security in the downlink of cellular networks, where each BS simultaneously transmits confidential messages to several users, and where the confidential messages transmitted to each user can be eavesdropped by both (i) other users in the same cell and (ii) users in other cells. Moreover, this paper takes into account the inter-cell interference generated by each BS, as well as the fact that malicious users can cooperate. This is a practical scenario that has not yet been addressed. In this paper, we model the locations of BSs and mobile users as two independent two-dimensional Poisson point processes (PPPs), and we analyze the performance of RCI precoding by combining tools from stochastic geometry and random matrix theory. Stochastic geometry (SG) is a powerful tool to study cellular networks with a random distribution of BSs and users \cite{Andrews11,DhillonJSAC}, whereas random matrix theory (RMT) enables a deterministic abstraction of the physical layer, for a fixed cellular network topology \cite{CouilletBook}. By combining results from SG and RMT, we can explicitly characterize the achievable secrecy rates accounting for (i) the spatial distribution of BSs and users, and (ii) the fluctuations of their channels. Our main contributions are summarized below.
\begin{itemize}
\item We obtain an approximation for the probability of secrecy outage $\hat{\mathcal{P}}_o$ in a cellular network under RCI precoding. We find regimes where RCI precoding achieves confidential communication with probability of secrecy outage $\hat{\mathcal{P}}_o < 1$. We also find that since cellular networks are interference-and-leakage-limited, they tend to be in secrecy outage w.p. 1 if the transmit power grows unbounded. This is different to the case of an isolated cell, where the probability of secrecy outage can be made arbitrarily small by increasing the number of transmit antennas \cite{GeraciTWC}.
\item We derive an approximation for the mean per-user secrecy rate achievable by RCI precoding in a cellular network. We find that RCI precoding can achieve a non-zero secrecy rate, however the secrecy rate in a cellular network does not grow unbounded with the transmit SNR. This is different to the case of an isolated cell, where an achievable secrecy rate can grow monotonically with the transmit SNR for a sufficient number of transmit antennas \cite{GeraciJSAC}.
\item We show that in a cellular network there is an optimal value for the density of BSs $\lambda_b$ that maximizes the mean secrecy rate. The value of $\lambda_b$ trades off useful signal power, interference, and information leakage. We find that the optimal deployment density $\lambda_b$ is a decreasing function of the SNR.
\end{itemize}

The remainder of the paper is organized as follows. Section II introduces the downlink of a cellular network with malicious users. In Section III, we characterize the secrecy rates achievable by RCI precoding. In Section IV, we derive approximations for the probability of secrecy outage and the mean secrecy rate. In Section V, we provide numerical results to confirm the accuracy of the analysis. The paper is concluded in Section VI.
\section{System Model}

\subsection{Network Topology}

\begin{figure}
\centering
\includegraphics[width=\figwidth]{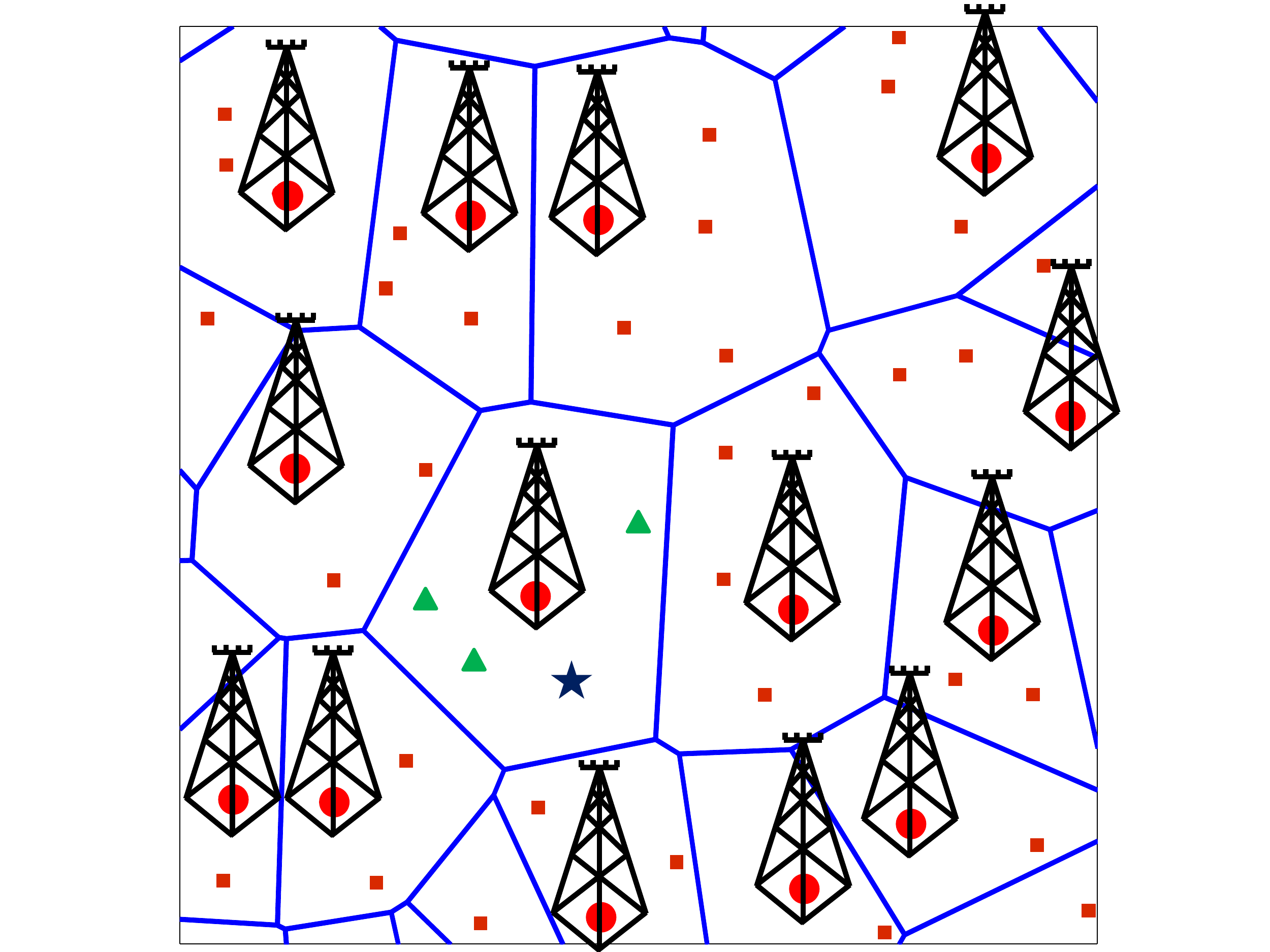}
\caption{Illustration of a cellular network. The circles, squares, and triangles denote BSs, out-of-cell users, and in-cell users, respectively. The star denotes a typical user as discussed in Subsection II-C.}
\label{fig:cellular}
\end{figure}

We consider the downlink of a cellular network, as depicted in Fig. \ref{fig:cellular}. Each BS transmits at power $P$ and is equipped with $N$ antennas. The locations of the BSs are drawn from a homogeneous PPP $\Phi_b$ of density $\lambda_b$. We consider single-antenna users, and assume that each user is connected to the closest BS. The locations of the users are drawn from an independent PPP $\Phi_u$ of density $\lambda_u$. We denote by $\mathcal{K}_b$ and by $K_b = |\mathcal{K}_b|$ the set of users and the number of users connected to the BS $b$, respectively. We denote by $\mathbf{H}_b = \left[\mathbf{h}_{b,1},\ldots,\mathbf{h}_{b,K_b} \right]^{\dagger}$ the $K_b \times N$ channel matrix for the BS $b$, where $\mathbf{h}_{b,j} \sim \mathcal{CN}(\mathbf{0},\mathbf{I})$ is the channel vector that accounts for the fading between the BS $b$ and the user $j \in \mathcal{K}_b$.

\subsection{RCI Precoding}
Transmission takes place over a block fading channel. The signal transmitted by the generic BS $b$ is $\mathbf{x}_b = \left[x_{b,1},\ldots,x_{b,N} \right]^{T} \in \mathbb{C}^{N \times 1}$. We consider RCI precoding because it is a linear scheme that allows low-complexity implementation \cite{Spencer04Magazine,Li10a}. Although suboptimal, RCI precoding is particularly interesting because it can control the amount of crosstalk between the users \cite{Joham05,Peel05,Ryan09}. In RCI precoding, the transmitted vector $\mathbf{x}_b$ is obtained at the BS $b$ by performing a linear processing on the vector of confidential messages $\mathbf{m}_b = \left[m_{b,1},\ldots,m_{b,K_b} \right]^{T}$, whose entries are chosen independently, satisfying $\mathbb{E}[ \left|m_{b,j}\right|^2 ] =1, \forall j$. The transmitted signal $\mathbf{x}_b$ after RCI precoding can be written as $\mathbf{x}_b = \sqrt{P}\mathbf{W}_b \mathbf{m}_b$, where $\mathbf{W}_b = \left[\mathbf{w}_{b,1},\ldots,\mathbf{w}_{b,K_b} \right]$ is the $N \times K_b$ RCI precoding matrix, given by \cite{Peel05}
\begin{equation}
\mathbf{W}_b = \frac{1}{\sqrt{\zeta_b}} \mathbf{H}_b^{\dagger} \left( \mathbf{H}_b \mathbf{H}_b^{\dagger} + N \xi \mathbf{I}_{K_b} \right) ^{-1}
\end{equation}
and $\zeta_b = \textrm{tr} \left\{ \mathbf{H}_b^{\dagger} \mathbf{H}_b \left( \mathbf{H}_b^{\dagger} \mathbf{H}_b + N \xi \mathbf{I}_N \right) ^{-2} \right\}$ is a long-term power normalization constant. The function of the real regularization parameter $\xi$ is to achieve a tradeoff between the signal power at the legitimate user and the crosstalk at the other users served by the same BS. The optimal value for the parameter $\xi$ in cellular networks is unknown, and we leave its calculation as a future work. Since the results obtained in this paper hold for any value of $\xi$, we will now assume that each BS sets $\xi$ to the value that maximizes the large-system secrecy rate in an isolated cell, given by \cite{GeraciJSAC}
\begin{equation}
\xi = \frac {-2\rho^2\left( 1 - \beta \right)^2 + 6\rho \beta + 2\beta^2 - 2 \left[ \beta \left( \rho+1 \right) -\rho \right] \cdot \sqrt{ \beta^2 \left[ \rho^2 + \rho + 1 \right] - \beta \left[ 2 \rho \left( \rho -1 \right) \right] + \rho^2 } } {6 \rho^2 \left( \beta + 2 \right) + 6 \rho \beta},
\label{eqn:xi_opt}
\end{equation}
where $\beta$ is the ratio between the number of users in the cell and the number of antennas at the BS.

\subsection{Malicious Users}

In general, the BSs cannot determine the behavior of the users, i.e., whether they act maliciously as eavesdroppers or not. As a worst-case scenario, we assume that for each legitimate user, all the remaining users in the network can act as eavesdroppers. For a user $o$ connected to the BS $b$, the set of $K_{b}-1$ malicious users within the same cell is denoted by $\mathcal{M}_o^I = \mathcal{K}_b \backslash o$, and the set formed by the rest of the malicious users in the network is denoted by $\mathcal{M}_o^E = \Phi_u \backslash \mathcal{K}_b$. In Fig. \ref{fig:cellular}, the legitimate user $o$, the set of (intra-cell) malicious users $\mathcal{M}_o^I$, and the set of (external) malicious users $\mathcal{M}_o^E$ are represented by star, triangles, and squares, respectively. The total set of malicious users for the legitimate receiver $o$ is denoted by $\mathcal{M}_o = \mathcal{M}_o^I \cup \mathcal{M}_o^E = \Phi_u \backslash o$. It is important to make such a distinction between the intra-cell malicious users in $\mathcal{M}_o^I$ and the external malicious users in $\mathcal{M}_o^E$. In fact, the BS $b$ knows the channels of the intra-cell malicious users in $\mathcal{M}_o^I \subset \mathcal{K}_b$, and exploits this information by choosing an RCI precoding matrix $\mathbf{W}_b$ which is a function of these channels. The RCI precoding thus controls the amount of information leakage at the malicious users in $\mathcal{M}_o^I$. On the other hand, the BS $b$ does not know the channels of all the other external malicious users in $\mathcal{M}_o^E$, and $\mathbf{W}_b$ does not depend upon these channels. Therefore, the signal received by the malicious users in $\mathcal{M}_o^E$ is not directly affected by RCI precoding.
\section{Achievable Secrecy Rates}

\begin{table}
\centering
\caption{Notation Summary}
\label{table:notationtable}
\begin{tabulary}{\columnwidth}{ |c | C | }
\hline
    \textbf{Notation} & \textbf{Description} \\ \hline
    $\Phi_b$; $\lambda_b$		&	A PPP modeling the locations of BSs; deployment density of BSs \\ \hline
    $\Phi_u$; $\lambda_u$ 		&	An independent PPP modeling the locations of users; density of the users \\ \hline
    $P$; $\rho$ & Downlink transmit power for each BS; SNR $\rho \triangleq \frac{P}{\sigma^2}$ \\ \hline
    $N$; $c$ 	& Number of transmit antennas for each BS; BS which is closest to the origin $o$ \\ \hline
    $\mathcal{K}_b$; $K_b=\left| \mathcal{K}_b \right|$ & Set of users associated with BS $b$; number of users associated with BS $b$ \\ \hline
    $\mathcal{M}_o^I = \mathcal{K}_c \backslash o$ & Set formed by the $K_c - 1$ malicious users in the same cell as the typical user \\ \hline
    $\mathcal{M}_o^E = \Phi_u \backslash \mathcal{K}_c$ & Set formed by the malicious users in all the other cells \\ \hline    
    $\mathbf{m}_b$; $\mathbf{x}_b = \sqrt{P} \mathbf{W}_b \mathbf{m}_b$ & Confidential messages sent by BS $b$ to its users; signal transmitted by BS $b$ \\ \hline
    $\mathbf{h}_{b,j} \sim \mathcal{CN}(\mathbf{0},\mathbf{I})$; $\eta$  & Channel vector between BS $b$ and user $j$; path loss exponent \\ \hline
    $g_{b,o} \sim \Gamma(K_b,1)$		&	Inter-cell interference power gain from BS $b$ to the typical user in $o$  \\ \hline
    $g_{c,e} \sim \textrm{exp}(1)$  &	Leakage power gain from BS $c$ to the malicious user $e \in \mathcal{M}_o^E$  \\ \hline
\end{tabulary}
\end{table}

In this section, we derive a secrecy rate achievable by RCI precoding for the typical user in the downlink of a cellular network.

\subsection{SINR at a Typical User}

We consider a typical user $o$ located at the origin, and connected to the closest BS, located in $c \in \Phi_b$. The distance between the typical user and the closest BS is given by $\|c\|$. The typical user receives self-interference caused by the other messages $m_{c,u}, u \neq o$ transmitted by the BS $c$, and inter-cell interference caused by the signal transmitted by all the other BSs $b \in \Phi_b \backslash c$. The signal received by the typical user is given by
\begin{equation}
y_{o} = \sqrt{P \, \|c\|^{-\eta}} \, \mathbf{h}_{c,o}^{\dagger} \mathbf{w}_{c,o} m_{c,o} + \sqrt{P \, \|c\|^{-\eta}} \sum_{u \in \mathcal{K}_c \backslash o} \! \mathbf{h}_{c,o}^{\dagger} \mathbf{w}_{c,u} m_{c,u} + \! \sum_{b \in \Phi_b \backslash c} \! \sqrt{P \, \|b\|^{-\eta}} \, 
\sum_{j=1}^{K_b} \mathbf{h}_{b,o}^{\dagger} \mathbf{w}_{b,j} m_{b,j} + n_o
\label{eqn:received_L}
\end{equation}
where $\|b\|$ is the distance between the typical user and the generic BS $b$, and $\eta$ is the path loss exponent. The four terms in (\ref{eqn:received_L}) represent the useful signal, the crosstalk (or self-interference), the inter-cell interference, and the thermal noise seen at the typical user, respectively. The latter is given by $n_o \sim \mathcal{CN}(0,\sigma^2)$, and we define the SNR as $\rho \triangleq P/ \sigma ^2$. 

We assume that the legitimate receiver at $o$ treats the interference power as noise. The SINR $\go$ at the legitimate receiver $o$ is given by
\begin{equation}
\go = \frac {\rho \|c\|^{-\eta} \left| \mathbf{h}_{c,o}^{\dagger} \mathbf{w}_{c,o} \right| ^2} {\rho \|c\|^{-\eta} \sum_{u \in \mathcal{K}_c \backslash o} \left| \mathbf{h}_{c,o}^{\dagger} \mathbf{w}_{c,u} \right|^2 + \rho \sum_{b \in \Phi_b \backslash c} \frac{g_{b,o}}{K_b} \|b\|^{-\eta} + 1},
\label{eqn:gammaL}
\end{equation}
where we define $\tilde{\mathbf{w}}_{b,j} \triangleq \sqrt{K_b} \mathbf{w}_{b,j}$ and
\begin{equation}
g_{b,o} \triangleq \sum_{j=1}^{K_b} \left|  \mathbf{h}_{b,o}^{\dagger} \tilde{\mathbf{w}}_{b,j} \right|^2.
\label{eqn:gbo}
\end{equation}

\subsection{SINR at the Malicious Users}

The cell where the typical user $o$ is located is referred to as the \textit{tagged cell}. For the typical user $o$, the set of malicious users is denoted by $\mathcal{M}_o = \mathcal{M}_o^I \cup \mathcal{M}_o^E$, where $\mathcal{M}_o^I = \mathcal{K}_c \backslash o$ is the set of remaining users in the tagged cell, and $\mathcal{M}_o^E = \Phi_u \backslash \mathcal{K}_c$ is the set of all users in other cells.

\begin{figure}[t]
\centering
\begin{tikzpicture}[scale=.6]
\coordinate (A1) at (8.1472,6.5574);
\coordinate (A4) at (9.1338,9.3399);
\coordinate (A5) at (6.3236,6.7874);
\coordinate (A7) at (2.7850,7.4313);
\coordinate (A8) at (5.4688,3.9223);
\coordinate (A9) at (9.5751,6.5548);
\coordinate (A10) at (1.6489,1.7119);
\coordinate (A11) at (1.5761,7.0605);
\coordinate (A13) at (9.5717,2.7692);
\coordinate (A14) at (4.8538,0.4617);
\coordinate (A15) at (3.0028,0.9713);    
\coordinate (A16) at (1.4189,8.2346);
\coordinate (A17) at (4.2176,6.9483);
\coordinate (A18) at (9.1574,3.1710);
\coordinate (A19) at (7.9221,9.5022);
\coordinate (A20) at (9.5949,0.3445); 
\coordinate (A21) at (4.3874,2.7603);
\coordinate (A22) at (3.8156,6.7970);
\coordinate (A23) at (7.6552,6.5510);
\coordinate (A24) at (7.9520,1.6261);
\coordinate (A25) at (1.8687,1.1900);    
\coordinate (A26) at (4.8976,4.9836);
\coordinate (A27) at (4.4559,9.5974);
\coordinate (A28) at (6.4631,3.4039);
\coordinate (A29) at (7.0936,5.8527);
\coordinate (A30) at (7.5469,2.2381);

\begin{scope}[fill opacity=0.5]
\draw [orange!80!,fill=orange!50!] (A1) circle (1.5); 
\draw [orange!80!,fill=orange!50!] (A4) circle (1.5); 
\draw [orange!80!,fill=orange!50!] (A5) circle (1.5);
\draw [orange!80!,fill=orange!50!] (A7) circle (1.5); 
\draw [orange!80!,fill=orange!50!] (A8) circle (1.5);
\draw [orange!80!,fill=orange!50!] (A9) circle (1.5);
\draw [orange!80!,fill=orange!50!] (A10) circle (1.5); 
\draw [orange!80!,fill=orange!50!] (A11) circle (1.5);
\draw [orange!80!,fill=orange!50!] (A13) circle (1.5); 
\draw [orange!80!,fill=orange!50!] (A14) circle (1.5);
\draw [orange!80!,fill=orange!50!] (A15) circle (1.5);
\draw [orange!80!,fill=orange!50!] (A16) circle (1.5); 
\draw [orange!80!,fill=orange!50!] (A17) circle (1.5);
\draw [orange!80!,fill=orange!50!] (A18) circle (1.5);
\draw [orange!80!,fill=orange!50!] (A19) circle (1.5); 
\draw [orange!80!,fill=orange!50!] (A20) circle (1.5);
\draw [orange!80!,fill=orange!50!] (A21) circle (1.5);
\draw [orange!80!,fill=orange!50!] (A22) circle (1.5);
\draw [orange!80!,fill=orange!50!] (A23) circle (1.5); 
\draw [orange!80!,fill=orange!50!] (A24) circle (1.5);
\draw [orange!80!,fill=orange!50!] (A25) circle (1.5);
\draw [orange!80!,fill=orange!50!] (A26) circle (1.5); 
\draw [orange!80!,fill=orange!50!] (A27) circle (1.5);
\draw [orange!80!,fill=orange!50!] (A28) circle (1.5);
\draw [orange!80!,fill=orange!50!] (A29) circle (1.5); 
\draw [orange!80!,fill=orange!50!] (A30) circle (1.5);
\end{scope}

\filldraw [black] (A1) circle (2pt);
\filldraw [black] (A4) circle (2pt);
\filldraw [black] (A5) circle (2pt);
\filldraw [black] (A7) circle (2pt);
\filldraw [black] (A8) circle (2pt);
\filldraw [black] (A9) circle (2pt);
\filldraw [black] (A10) circle (2pt);
\filldraw [black] (A11) circle (2pt);
\filldraw [black] (A13) circle (2pt);
\filldraw [black] (A14) circle (2pt);
\filldraw [black] (A15) circle (2pt);
\filldraw [black] (A16) circle (2pt);
\filldraw [black] (A17) circle (2pt);
\filldraw [black] (A18) circle (2pt);
\filldraw [black] (A19) circle (2pt);
\filldraw [black] (A20) circle (2pt);
\filldraw [black] (A21) circle (2pt);
\filldraw [black] (A22) circle (2pt);
\filldraw [black] (A23) circle (2pt);
\filldraw [black] (A24) circle (2pt);
\filldraw [black] (A25) circle (2pt);
\filldraw [black] (A26) circle (2pt);
\filldraw [black] (A27) circle (2pt);
\filldraw [black] (A28) circle (2pt);
\filldraw [black] (A29) circle (2pt);
\filldraw [black] (A30) circle (2pt);

\end{tikzpicture}
\caption{Example of percolation in a random plane network. Dots represent malicious users, and discs represents the cooperation range of malicious users. Two malicious users can cooperate when their respective discs overlap.}
\label{fig:percolation}
\end{figure}
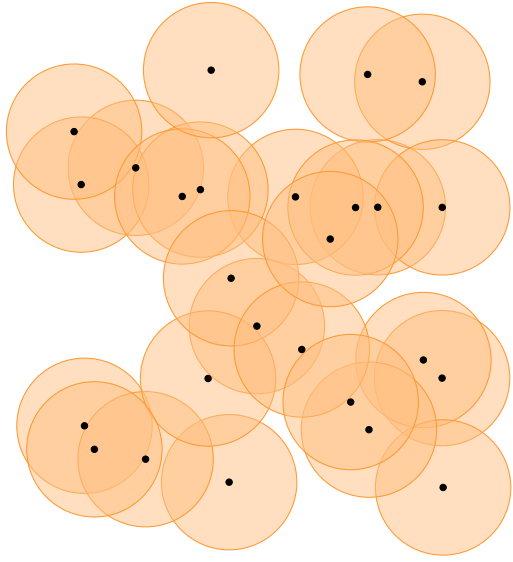

We assume that each malicious user can communicate directly to any other malicious user within a cooperation radius $r_c$ around it, i.e., cooperation is possible for distances smaller than $r_c$. This assumption comes from the following model: assume that malicious users can transmit at a certain power $P_M$ and that their signal is attenuated over distance according to a deterministic decreasing function $l(d)$. Assume also that malicious users can succesfully receive data if the signal is at least $t$ times stronger than the ambient noise, which has power $P_n$. Even under the condition that the interference is perfectly canceled, there is a maximum distance beyond which the two users cannot cooperate. This distance, referred to as the cooperation radius, is given by
\begin{equation}
r_c \triangleq \max \left\{ d: \frac{P_M \, l(d)}{P_n} \geq t \right\}.
\end{equation}
By connecting each pair of cooperating malicious users, it is possible to generate a random plane network \cite{Gilbert61}, which represents an infinite cooperation network of malicious users with range $r_c$. It is known that if the density of users satisfies $\lambda_u > \frac{8 \log 2}{r_c^2}$, the random plane network contains an infinite cluster almost surely. An illustration of this phenomenon, known as percolation, is provided in Fig. \ref{fig:percolation}. As a result, if the density of users $\lambda_u$ is large enough, then percolation will occur, and there will be an infinite set of malicious users cooperating to eavesdrop the message intended for the typical user.


Motivated by these observations, in the following we will consider the worst-case scenario where all the malicious users in $\mathcal{M}_o$ can cooperate to eavesdrop on the message intended for the typical user in $o$. Since each malicious user is likely to decode its own message, it can indirectly pass this information to all the other malicious users. In the worst-case scenario, all the malicious users in $\mathcal{M}_o$ can therefore subtract the interference generated by all the messages $m_j$, $j\neq o$.

After interference cancellation, the signal received at a malicious user $i \in\mathcal{M}_o^I$ in the tagged cell is given by
\begin{equation}
y_{i} = \sqrt{P \, \|i-c\|^{-\eta}} \, \mathbf{h}_{c,i}^{\dagger} \mathbf{w}_{c,o} m_{c,o} + n_i
\label{eqn:received_MI}
\end{equation}
where $\|i-c\|$ is the distance between the BS $c$ and the malicious user $i \in \mathcal{M}_o^I$. The signal received at a malicious user $e \in\mathcal{M}_o^E$ outside the tagged cell is given by
\begin{equation}
y_{e} = \sqrt{P \, \|e-c\|^{-\eta}} \, \mathbf{h}_{c,e}^{\dagger} \mathbf{w}_{c,o} m_{c,o} + n_e.
\label{eqn:received_ME}
\end{equation}
We denote by $\gamma_i$ and $\gamma_e$ the SINRs at the malicious users $i \in \mathcal{M}_o^I$ and $e \in \mathcal{M}_o^E$, respectively.

Due to the cooperation among all malicious users in $\mathcal{M}_o = \mathcal{M}_o^I \cup \mathcal{M}_o^E$, the set $\mathcal{M}_o$ can be seen as a single equivalent multi-antenna malicious user, denoted by $M_o$. After interference cancellation, $M_o$ sees the useful signal embedded in noise, therefore applying maximal ratio combining is optimal, and yields to an SINR given by
\begin{equation}
\gMo = \sum_{i \in\mathcal{M}_o^I} \gamma_i + \sum_{e \in\mathcal{M}_o^E} \gamma_e = \rho \, \sum_{i \in\mathcal{M}_o^I} { \|i-c\|^{-\eta}  \left|\mathbf{h}_{c,i}^{\dagger} \mathbf{w}_{c,o}\right|^2} + \frac{\rho}{K_c} \sum_{e \in\mathcal{M}_o^E} {g_{c,e} \|e-c\|^{-\eta}},
\label{eqn:gammaM}
\end{equation}
where
\begin{equation}
g_{c,e} \triangleq \left|  \mathbf{h}_{c,e}^{\dagger} \tilde{\mathbf{w}}_{c,o} \right|^2,
\label{eqn:gce}
\end{equation}
with $\tilde{\mathbf{w}}_{c,o} \triangleq \sqrt{K_c} \mathbf{w}_{c,o}$ and $n_i, n_e \sim \mathcal{CN}(0,\sigma^2)$.

\subsection{Achievable Secrecy Rates}

We are now able to obtain an expression for the secrecy rate achievable by RCI precoding for the typical user of a downlink cellular network.

\begin{Proposition}
A secrecy rate achievable by RCI precoding for the typical user $o$ is given by
\begin{align}
R &\triangleq \Bigg\{ \log_2 \Bigg( 1 + \frac {\rho \|c\|^{-\eta} \left| \mathbf{h}_{c,o}^{\dagger} \mathbf{w}_{c,o} \right| ^2} {\rho \|c\|^{-\eta} \sum_{i \in \mathcal{M}_o^I} \left| \mathbf{h}_{c,o}^{\dagger} \mathbf{w}_{c,i} \right|^2 + \rho I + 1 } \Bigg) \nonumber\\
& \quad \quad \quad - \log_2 \Bigg( 1 + \rho \, \sum_{i \in\mathcal{M}_o^I} {\|i-c\|^{-\eta} \left|\mathbf{h}_{c,i}^{\dagger} \mathbf{w}_{c,o}\right|^2} + \rho L \Bigg) \Bigg\}^+,
\label{eqn:R_complete}
\end{align}
where we use the notation $\left\{x\right\}^+ \triangleq \max \left(x,0\right)$, and where $I$ and $L$ represent the interference and leakage term, respectively, given by
\begin{align}
I &= \sum_{b \in \Phi_b \backslash c} \frac{g_{b,o}}{K_b} \|b\|^{-\eta}
\label{eqn:I} \\
L &= \frac{1}{K_c} \sum_{e \in \mathcal{M}_o^E} {g_{c,e} \|e-c\|^{-\eta} }.
\label{eqn:L}
\end{align}
\end{Proposition}
\begin{IEEEproof}
The BS $c$, the user $o$, and the equivalent malicious user $M_o$ form
an equivalent multi-input, single-output, multi-eavesdropper
(MISOME) wiretap channel \cite{Khisti10I}. As a result, an achievable secrecy rate is given by \cite{Geraci12}
\begin{equation}
R = \left\{ \log_2 \left( 1 + \go \right) - \log_2 \left( 1 + \gMo \right) \right\}^+.
\label{eqn:R_simple}
\end{equation}
Substituting (\ref{eqn:gammaL}) and (\ref{eqn:gammaM}) in (\ref{eqn:R_simple}) yields (\ref{eqn:R_complete}).
\end{IEEEproof}

The statistics of the terms $g_{b,o}$ and $g_{c,e}$ in (\ref{eqn:gbo}) and (\ref{eqn:gce}), respectively, can be characterized as follows \cite{DhillonMIMO}.
\begin{Proposition}
For regularized channel inversion precoding we have that (i) the inter-cell interference power gain at the typical legitimate user $o$ is distributed as $g_{b,o} \sim \Gamma(K_b,1)$, and (ii) the leakage power gain at the malicious user $e \in \mathcal{M}_o^E$ is distributed as $g_{c,e} \sim \mathrm{exp}(1)$.
\label{Proposition:distributions}
\end{Proposition}
\begin{IEEEproof}
See Appendix A.
\end{IEEEproof}

We now define the probability of secrecy outage and the mean secrecy rate for the typical user.
\begin{Definition}
The probability of secrecy outage for the typical user $o$ is defined as
\begin{equation}
\mathcal{P}_o \triangleq \mathbb{P} (R \leq 0).
\label{eqn:outage_definition}
\end{equation}
\end{Definition}
The probability of secrecy outage also denotes the fraction of time for which a BS cannot transmit to a typical user at a non-zero secrecy rate.

\begin{Definition}
The mean secrecy rate for the typical user $o$ is defined as
\begin{equation}
R_m \triangleq \mathbb{E} \left[ R \right].
\label{eqn:mean_rate_definition}
\end{equation}
\end{Definition}
\section{Large-system Analysis}

In this section, we derive approximations for (i) the secrecy outage probability, i.e., the probability that the secrecy rate $R$ achievable by RCI precoding for the typical user $o$ is zero, and (ii) the mean secrecy rate achievable by RCI precoding in the downlink of a cellular network.

\subsection{Preliminaries}

Throughout the analysis, we make the following assumptions.

\begin{Assumption}
For uniformity of notation, we assume $K_c = K_b = K \triangleq \frac{\lambda_u}{\lambda_b}$, $\forall b$, i.e., we approximate the number of users served by each BS by its average value, given by the ratio between the density of users and the density of BSs. In order for this equivalence to hold, we ignore a small bias that makes the tagged cell bigger than a typical cell. This bias is a result of Feller's paradox, also known as waiting bus paradox in one dimension \cite{BacBreB2003}.
\end{Assumption}

\begin{Assumption}
We assume $\|i-c\|\approx\|c\|$, $\forall i \in \mathcal{M}_o^I$, i.e., we approximate the distance between the tagged BS $c$ and each user connected to $c$ by the distance between the BS $c$ and the typical user $o$. We then approximate the Voronoi region of the tagged BS $c$ by a ball centered at $c$ and with radius $r = \frac{1}{\sqrt{\pi \lambda_b}}$, i.e., $\mathcal{B}(c,r) \triangleq \left\{ m \in \mathbb{R}^2, \|m-c\| \leq r \right\}$. For the sake of consistency, the value of $r$ is chosen to ensure that $\mathcal{B}(c,r)$ has the same area as the average cell.
\end{Assumption}

Note that despite these assumptions, which are necessary to maintain tractability, our analysis captures all the key characteristics of the cellular networks that affect physical layer security, as discussed in the sequel. The simplified model also provides some fundamental insights into the dependence of key performance metrics, such as secrecy outage and mean secrecy rate, on the transmit power and BS deployment density.

Under Assumptions 1 and 2, we obtain the approximations $\mathcal{M}_o^I \approx \hat{\mathcal{M}}_o^I$ and $\mathcal{M}_o^E \approx \hat{\mathcal{M}}_o^E$, where $\hat{\mathcal{M}}_o^I$ is a set of $K-1$ malicious users located at distance $\|c\|$ from the BS $c$, and $\hat{\mathcal{M}}_o^E$ is a set given by 
\begin{equation}
\hat{\mathcal{M}}_o^E = \{ e \in \Phi_u \cap \bar{\mathcal{B}}(c,r) \},
\end{equation}
with $\bar{\mathcal{B}}$ denoting the complement of the set $\mathcal{B}$. We can then approximate the interference and leakage terms in (\ref{eqn:I}) and (\ref{eqn:L}) as follows
\begin{align}
I &\approx \hat{I} = \frac{1}{K} \sum_{b \in \Phi_b \backslash c} g_{b,o} \|b\|^{-\eta}
\label{eqn:I_tilde}\\
L &\approx \hat{L} = \frac{1}{K} \sum_{e \in \hat{\mathcal{M}}_o^E} {g_{c,e} \|e-c\|^{-\eta}}.
\label{eqn:L_tilde}
\end{align}

We now carry out a large-system analysis by assuming that both (i) the average number of users $K$ in each cell, and (ii) the number of transmit antennas $N$ at the each BS grow to infinity in a fixed ratio $\beta \triangleq \frac{K}{N}$. We can thus approximate the remaining random quantities in (\ref{eqn:R_complete}) by their large-system deterministic equivalents \cite{NguyenGCOM08,Wagner12}
\begin{equation}
\left| \mathbf{h}_{c,o}^{\dagger} \mathbf{w}_{c,o} \right| ^2 \approx \alpha, \quad \sum_{i \in \hat{\mathcal{M}}_o^I} \left| \mathbf{h}_{c,o}^{\dagger} \mathbf{w}_{c,i} \right|^2 \approx \chi, \quad \textrm{and} \quad \sum_{i \in \hat{\mathcal{M}}_o^I} {\left|\mathbf{h}_{c,i}^{\dagger} \mathbf{w}_{c,o}\right|^2} \approx \chi,
\end{equation}
where
\begin{equation}
\alpha = \frac {g\left(\beta,\xi\right)\left\{1+\frac{\xi}{\beta}\left[1+g\left(\beta,\xi\right)\right]^2\right\}}{\left[1+g\left(\beta,\xi\right)\right]^2}, \quad \chi = \frac {1}{\left[1+g\left(\beta,\xi\right)\right]^2},
\end{equation}
and
\begin{equation}
g \left( \beta,\xi \right) = \frac{1}{2} \left[ \sqrt{ \frac{\left(1-\beta \right)^2}{\xi^2}  +  \frac{2\left(1+\beta\right)}{\xi}  +  1} +  \frac{1-\beta}{\xi}  -  1 \right],
\end{equation}
and where it follows from (\ref{eqn:xi_opt}) that
\begin{equation}
\lim_{\rho \rightarrow \infty} \chi = 0, \enspace	\textrm{for} \enspace \beta \leq 1.
\label{eqn:chi}
\end{equation}

An approximated secrecy rate can be therefore obtained as follows.
\begin{Definition}
An approximation for the achievable secrecy rate $R$ is given by
\begin{equation}
R \approx \hat{R} \triangleq \Bigg\{ \log_2 \Bigg( 1 + \frac {\rho \alpha \|c\|^{-\eta}} {\rho \chi \|c\|^{-\eta} + \rho \hat{I} + 1 } \Bigg) - \log_2 \Bigg( 1 + \rho \, \chi \|c\|^{-\eta} + \rho \hat{L} \Bigg) \Bigg\}^+.
\label{eqn:R_hat}
\end{equation}
\end{Definition}

\begin{figure}
\centering
\includegraphics[width=\figwidth]{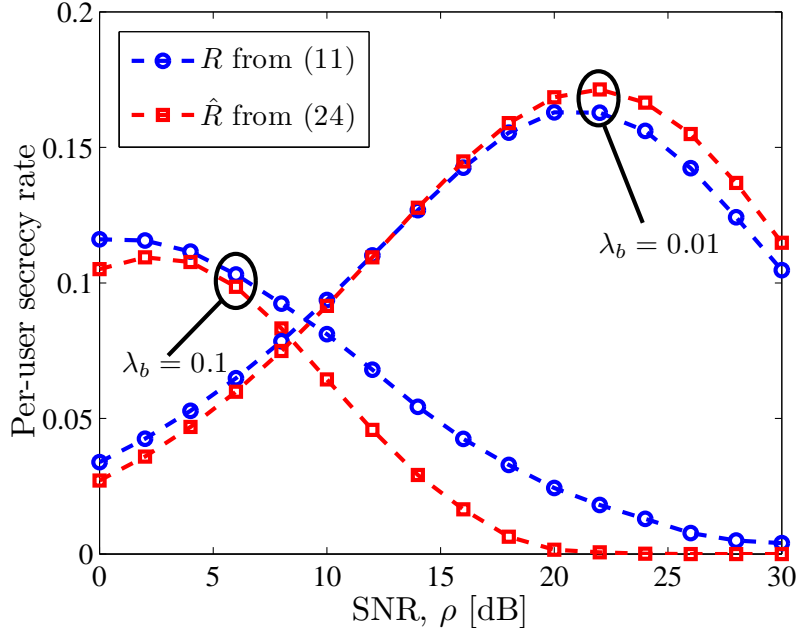}
\caption{Comparison between the simulated ergodic secrecy rate $R$ in (\ref{eqn:R_complete}) and the approximation $\hat{R}$ in (\ref{eqn:R_hat}) versus the SNR, for $N=20$ transmit antennas, an average of $K=20$ users per BS, and $\eta=4$.}
\label{fig:R_vs_Rhat}
\end{figure}

In Fig. \ref{fig:R_vs_Rhat} we compare the simulated ergodic secrecy rate $R$ in (\ref{eqn:R_complete}) to the approximation $\hat{R}$ in (\ref{eqn:R_hat}), obtained in the large-system regime under Assumptions 1 and 2. The secrecy rates $R$ and $\hat{R}$ are plotted versus the SNR $\rho$, for a system with $N=20$ transmit antennas, an average number $K=20$ of users per cell, a path loss exponent $\eta=4$, and two values of the density of BS $\lambda_b$. 
Fig. \ref{fig:R_vs_Rhat} shows that $R$ and $\hat{R}$ follow the same trend, and that the approximation $R \approx \hat{R}$ is reasonable. The figure also shows that in a cellular network the secrecy rate does not monotonically increase with the SNR. A more detailed discussion on this phenomenon will be provided in Subsection IV-D.

\subsection{Characterization of $\hat{I}$ and $\hat{L}$}

We now provide some results on the Laplace transforms of the terms $\hat{I}$ and $\hat{L}$ which will be useful in the remainder of the paper.

\begin{Lemma}
The Laplace transform of the interference term $\mathcal{L}_{\hat{I}}(s,\|c\|) = \mathbb{E} [e^{-s\hat{I}}]$ is
\begin{equation}
\mathcal{L}_{\hat{I}}(s,\|c\|) = \mathrm{exp} \left( - \left(\frac{s}{K}\right)^{\frac{2}{\eta}} \lambda_b  C_{\eta,K}\left(s,\|c\|\right) \right)
\label{eqn:LaplaceI}
\end{equation}
where
\begin{equation}
C_{\eta,K}\left(s,\|c\|\right) \!=\! \frac{2 \pi}{\eta} \sum_{n=1}^{K} \! \binom{K}{n} \! \left[ B\left(1;K \!-\! n \!+\! \frac{2}{\eta},n\!-\!\frac{2}{\eta}\right) \!-\! B\left((1\!+\!\frac{s P}{K}\|c\|^{-\eta})^{-1};K \!-\! n \!+\! \frac{2}{\eta},n \!-\! \frac{2}{\eta}\right) \right],	
\end{equation}
and $B(x;y,z) = \int_0^x t^{y-1}(1-t)^{z-1} \mathrm{d} t$ is the incomplete Beta function.
\label{Lemma:LaplaceI}
\end{Lemma}
\begin{IEEEproof}
See Appendix B.
\end{IEEEproof}

\begin{Lemma}
The Laplace transform of the leakage term $\mathcal{L}_{\hat{L}}(s) = \mathbb{E} [e^{-s\hat{L}}]$ is
\begin{equation}
\mathcal{L}_{\hat{L}}(s) = \mathrm{exp} \left( - \lambda_u \left(\frac{s}{K}\right)^{\frac{2}{\eta}} D_{\eta}(s) \right)
\label{eqn:LaplaceL}
\end{equation}
where
\begin{equation}
D_{\eta}(s) = \frac{2 \pi}{\eta} \left[  B\left(1; \frac{2}{\eta},1-\frac{2}{\eta}\right) - B\left(\frac{1}{1+\frac{s}{K}r^{-\eta}}; \frac{2}{\eta},1-\frac{2}{\eta}\right) \right].	
\end{equation}
\label{Lemma:LaplaceL}
\end{Lemma}
\begin{IEEEproof}
The proof is omitted since it is similar to the proof of Lemma \ref{Lemma:LaplaceI}.
\end{IEEEproof}

The probability density functions (pdfs) $f_{\hat{I}}$ and $f_{\hat{L}}$ of $\hat{I}$ and $\hat{L}$, respectively, can be obtained by inverting the respective Laplace transforms $\mathcal{L}_{\hat{I}}$ and $\mathcal{L}_{\hat{L}}$. We now propose simple approximations for $f_{\hat{I}}$ and $f_{\hat{L}}$, using the following well-known results \cite{BaccelliBook1}.

\begin{Proposition}
The mean and the variance of the interference term $\hat{I}$ are respectively given by
\begin{align}
\mu_{\hat{I}} &= \frac{2 \pi \lambda_b \|c\|^{-(\eta-2)}}{\eta-2},
\label{eqn:mu_I}\\
\sigma^2_{\hat{I}} &= \frac{\pi \lambda_b \left( K + K^2 \right) \|c\|^{-2(\eta-1)}}{K^2 \left(\eta-1\right)},
\label{eqn:var_I}
\end{align}
whereas the mean and the variance of the leakage term $\hat{L}$ are respectively given by
\begin{align}
\mu_{\hat{L}} &= \frac{2 \pi \lambda_u r^{-(\eta-2)}}{K (\eta-2)},
\label{eqn:mu_L}\\
\sigma^2_{\hat{L}} &= \frac{2 \pi \lambda_u r^{-2(\eta-1)}}{K^2 \left(\eta-1\right)}.
\label{eqn:var_L}
\end{align}
\label{Proposition:lognormal}
\end{Proposition}
\begin{IEEEproof}
See Appendix C.
\end{IEEEproof}

We then approximate the pdfs of $\hat{I}$ and $\hat{L}$ by lognormal distributions with the same respective mean and variance, as follows.
\begin{Definition}
The probability density functions of $\hat{I}$ and $\hat{L}$ can be approximated as follows
\begin{align}
f_{\hat{I}}(x) &\approx \frac{1}{x \sigma_{\hat{I},N}\sqrt{2 \pi}} \mathrm{exp} \left(-\frac{\left(\log x - \mu_{\hat{I},N} \right)^2}{2 \sigma^2_{\hat{I},N}}\right), \enspace x>0
\label{eqn:f_I_approx}\\
f_{\hat{L}}(z) &\approx \frac{1}{z \sigma_{\hat{L},N}\sqrt{2 \pi}} \mathrm{exp} \left(-\frac{\left(\log z - \mu_{\hat{L},N} \right)^2}{2 \sigma^2_{\hat{L},N}}\right), \enspace z>0
\label{eqn:f_L_approx}
\end{align}
where
\begin{align}
\mu_{\hat{I},N} &= \log \mu_{\hat{I}} - \frac{1}{2} \log \left(1+\frac{\sigma^2_{\hat{I}}}{\mu_{\hat{I}}^2}\right), \quad \sigma^2_{\hat{I},N} = \log \left(1+\frac{\sigma^2_{\hat{I}}}{\mu_{\hat{I}}^2}\right) \\
\mu_{\hat{L},N} &= \log \mu_{\hat{L}} - \frac{1}{2} \log \left(1+\frac{\sigma^2_{\hat{L}}}{\mu_{\hat{L}}^2}\right), \quad \sigma^2_{\hat{L},N} = \log \left(1+\frac{\sigma^2_{\hat{L}}}{\mu_{\hat{L}}^2}\right).
\end{align}
\label{D:lognormal}
\end{Definition}

\begin{figure}
\centering
\includegraphics[width=\figwidth]{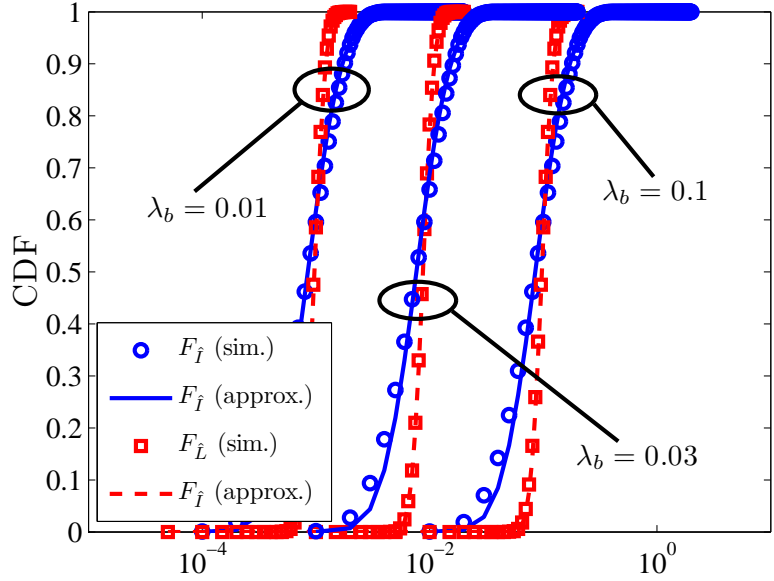}
\caption{Comparison between the simulated cumulative distribution functions (CDFs) of $\hat{I}$ and $\hat{L}$ and the lognormal approximations in (\ref{eqn:f_I_approx}) and (\ref{eqn:f_L_approx}), for an SNR $\rho=10$dB, $N=20$ transmit antennas, $K=20$ users per BS, $\|c\|=r$, and $\eta=4$.}
\label{fig:lognormal}
\end{figure}

In Fig. \ref{fig:lognormal} we compare the simulated cumulative distribution functions (CDFs) of $\hat{I}$ and $\hat{L}$ to the lognormal approximations provided in (\ref{eqn:f_I_approx}) and (\ref{eqn:f_L_approx}). The CDFs are plotted for an SNR $\rho=10$dB, $N=20$ transmit antennas, an average of $K=20$ users per BS, $\|c\|=r$, $\eta=4$, and three values of the density of BS $\lambda_b$. Fig. \ref{fig:lognormal} shows that the lognormal approximations provided in Definition \ref{D:lognormal} are accurate for all values of $\lambda_b$.

\subsection{Probability of Secrecy Outage}

We now obtain an approximation for the probability of secrecy outage with RCI precoding.
\begin{Theorem}
The probability of secrecy outage with RCI precoding can be approximated as
\begin{equation}
\mathcal{P}_o \approx \hat{\mathcal{P}}_o = \int_{0}^{\infty} \int_{-\infty}^{\infty} \int_{-\infty}^{\infty} \indic_{(z \geq \tau(x,y))} \, f_{\hat{L}}(z) \, \mathrm{d} z \, f_{\hat{I}}(x,y) \, \mathrm{d} x \, 2 \lambda_b \pi y e^{-\lambda_b \pi y^2} \, \mathrm{d} y,
\label{eqn:outage_generic}
\end{equation}
where
$f_{\hat{I}}(x,y)$ is the probability density function of the interference $\hat{I}$ for $\|c\|=y$, $f_{\hat{L}}(z)$ is the probability density function of the leakage $\hat{L}$, and where we have defined 
\begin{equation}
\tau(x,y) \triangleq	\frac {\alpha y^{-\eta}} {\rho \chi y^{-\eta} + \rho x + 1 } - \chi y^{-\eta}.
\label{eqn:tau}
\end{equation}
\label{T:outage_generic}
\end{Theorem}
\begin{IEEEproof}
By using approximation (\ref{eqn:R_hat}) in (\ref{eqn:outage_definition}), we obtain
\begin{align}
\mathcal{P}_o & \approx \mathbb{P} (\hat{R} \leq 0) = \mathbb{P} \left( \rho \, \chi \|c\|^{-\eta} + \rho \hat{L} \geq \frac {\alpha \|c\|^{-\eta}} {\chi \|c\|^{-\eta} + \hat{I} + \frac{1}{\rho} } \right) \nonumber\\
& = \mathbb{P} \left( \hat{L} \geq \frac {\alpha \|c\|^{-\eta}} {\rho \chi \|c\|^{-\eta} + \rho \hat{I} + 1 } - \chi \|c\|^{-\eta} \right) \nonumber\\
& \stackrel{(a)}{=} \int_{-\infty}^{\infty} \int_{-\infty}^{\infty} \mathbb{P} \left( \hat{L} \geq \tau(x,y) \right) f_{\hat{I}}(x,y \, | \, \|c\|=y) \, f_{\|c\|}(y) \, \mathrm{d} x \, \mathrm{d} y \nonumber \\
& = \int_{0}^{\infty} \int_{-\infty}^{\infty} \mathbb{E} \left[ \indic_{(\hat{L} \geq \tau(x,y))} \right] f_{\hat{I}}(x,y) \, \mathrm{d} x \, 2 \lambda_b \pi y e^{-\lambda_b \pi y^2} \, \mathrm{d} y \nonumber\\
& = \int_{0}^{\infty} \int_{-\infty}^{\infty} \int_{-\infty}^{\infty} \indic_{(z \geq \tau(x,y))} \, f_{\hat{L}}(z) \, \mathrm{d} z \, f_{\hat{I}}(x,y) \, \mathrm{d} x \, 2 \lambda_b \pi y e^{-\lambda_b \pi y^2} \, \mathrm{d} y,
\end{align}
where $(a)$ holds by defining $\tau(x,y)$ as in (\ref{eqn:tau}), and by noting that the distance $\|c\|$ between the typical user and the nearest BS $c$ has distribution \cite{Haenggi05}
\begin{equation}
f_{\|c\|}(y) = 2 \lambda_b \pi y \, \textrm{exp}(-\lambda_b \pi y^2), \enspace y>0.
\label{eqn:fc}
\end{equation}
\end{IEEEproof}

The result provided in Theorem 1 allows to evaluate the probability of secrecy outage without the need for Monte-Carlo simulations, which can be computationally expensive to account for all users and all exact Voronoi cells. Moreover, Theorem 1 yields to the following asymptotic result without the need to solve the integral.

In an isolated cell, a sufficient number of transmit antennas allows the BS to cancel the intra-cell interference and leakage, and to drive the probability of secrecy outage to zero \cite{GeraciTWC}. In a cellular network, the secrecy outage is also caused by the inter-cell interference and leakage, which cannot be controlled by the BS. It is easy to show that $\lim_{\rho \rightarrow \infty} \tau(x,y) \leq 0$, which from Theorem 1 implies $\lim_{\rho \rightarrow \infty} \hat{\mathcal{P}}_o = 1$. We therefore have the following result.

\begin{Remark}
In cellular networks, RCI precoding can achieve confidential communication with probability of secrecy outage $\hat{\mathcal{P}}_o < 1$. However unlike an isolated cell, cellular networks tend to be in secrecy outage w.p. $1$ if the transmit power grows unbounded, irrespective of the number of transmit antennas.
\label{R:outage}
\end{Remark}

\subsection{Mean Secrecy Rate}

In the following, we derive the mean secrecy rate achievable by RCI precoding.
\begin{Theorem}
The mean secrecy rate achievable by RCI precoding can be approximated as
\begin{align}
R_m \approx \hat{R}_m &= \int_0^{\infty} \int_{-\infty}^{\frac{\alpha}{\rho \chi}-\frac{1}{\rho} - \chi y^{-\eta}} \left\{ \log_2 \Bigg( 1 + \frac {\rho \alpha y^{-\eta}} {\rho \chi y^{-\eta} + \rho x + 1 } \Bigg) \int_{-\infty}^{\tau (x,y)} f_{\hat{L}}(z) \right. \nonumber \\
& \quad \left. - \int_{-\infty}^{\tau (x,y)} \log_2 \left( 1 + \rho \chi y^{-\eta} + \rho z \right) \, f_{\hat{L}}(z) \, \mathrm{d} z \right\} \, f_{\hat{I}}(x,y) \, \mathrm{d} x \, 2 \lambda_b \pi y e^{-\lambda_b \pi y^2} \, \mathrm{d} y.
\label{eqn:mean_rate_generic}
\end{align}
\label{T:mean_rate}
\end{Theorem}
\begin{IEEEproof}
See Appendix D.
\end{IEEEproof}

The result provided in Theorem 2 allows to evaluate the mean secrecy rate without the need for computationally expensive Monte-Carlo simulations. Moreover, Theorem 2 yields to the following asymptotic result without the need to solve the integral.

In an isolated cell, a sufficient number of transmit antennas allows the BS to cancel the intra-cell interference and leakage, and the secrecy rate increases monotonically with the SNR \cite{GeraciJSAC}. In a cellular network, the secrecy rate is also affected by the inter-cell interference and leakage, which cannot be controlled by the BS. It is easy to show that $\lim_{\rho \rightarrow \infty} \frac{\alpha}{\rho \chi} - \frac{1}{\rho}-\chi y^{-\eta} \leq 0$, which from Theorem 2 implies $\lim_{\rho \rightarrow \infty} \hat{R}_m = 0$. We therefore have the following result.

\begin{Remark}
In cellular networks, RCI precoding can achieve a non-zero secrecy rate $\hat{R}_m$. However unlike an isolated cell, the secrecy rate in a cellular network is interference-and-leakage-limited, and it cannot grow unbounded with the SNR, irrespective of the number of transmit antennas.
\label{R:SNR}
\end{Remark}

Theorem 2 shows that an optimal value for the BS deployment density $\lambda_b$ should be found as a tradeoff between (i) increasing the useful power $\alpha y^{-\eta}$, and (ii) reducing the intra-cell interference $\chi y^{-\eta}$ and leakage $\chi y^{-\eta}$, and the inter-cell interference $x$ and leakage $z$. We know from (\ref{eqn:chi}) that $\chi$ vanishes at high SNR, thus the terms $x$ and $z$ become dominant in (\ref{eqn:mean_rate_generic}). For a given cell load $K=\frac{\lambda_u}{\lambda_b}$, the terms $x$ and $z$ are minimized by small densities $\lambda_b$. We therefore have the following result which we will validate by simulations in Section V.

\begin{Remark}
In a cellular network with a fixed load, i.e., average number of users per BS, there is an optimal value for the deployment density of BSs that maximizes the mean secrecy rate, and this value is a decreasing function of the SNR. The optimal value of $\lambda_b$ can be found from (\ref{eqn:mean_rate_generic}) by performing a linear search. 
\label{R:lambda}
\end{Remark}

In order to calculate the mean secrecy rate in (\ref{eqn:mean_rate_generic}), one must obtain expressions for $f_{\hat{I}}$ and $f_{\hat{L}}$ via Laplace anti-transform or via approximations, as discussed in Section IV-B. We now derive a lower bound on $\hat{R}_m$ which can be calculated without knowledge of $f_{\hat{I}}$ and $f_{\hat{L}}$.

\begin{Corollary}
The approximated mean secrecy rate $\hat{R}_m$ can be lower bounded as
\begin{equation}
\hat{R}_m \geq \hat{R}_m^{LB} = \Bigg\{ \int_0^{\infty} \int_{-\infty}^{\infty} \Bigg[ \mathcal{F}^*_1(\phi,y) \mathcal{L}_{\hat{I}}(-i 2 \pi \phi,y) - \mathcal{F}^*_2(\phi,y) \mathcal{L}_{\hat{L}}(-i 2 \pi \phi) \Bigg] \mathrm{d} \phi \, 2 \lambda_b \pi y e^{- \lambda_b \pi y^2} \mathrm{d} y \Bigg\}^+,
\label{eqn:mean_rate_LB}
\end{equation}
with
\begin{align}
\mathcal{F}_1(\phi,y) &= \frac{\mathrm{sgn}(\phi) \, e^{2 \pi i (\chi y^{-\eta} + \frac{1}{\rho}) \phi}}{2 \phi \log 2}\left(1 - e^{2 \pi i \alpha \phi y^{-\eta}} \right),\\
\mathcal{F}_2(\phi,y) &= \frac{-e^{2 \pi i (\chi y^{-\eta} + \frac{1}{\rho}) \phi}}{\log 2} \left[ \frac{1}{2\left|\phi\right|} + \frac{\gamma}{\rho} \delta(\phi) \right].
\end{align}
\label{C:Plancherel}
\end{Corollary}
\begin{IEEEproof}
See Appendix E.
\end{IEEEproof}
\section{Numerical Results}

\begin{figure}
\centering
\includegraphics[width=\figwidth]{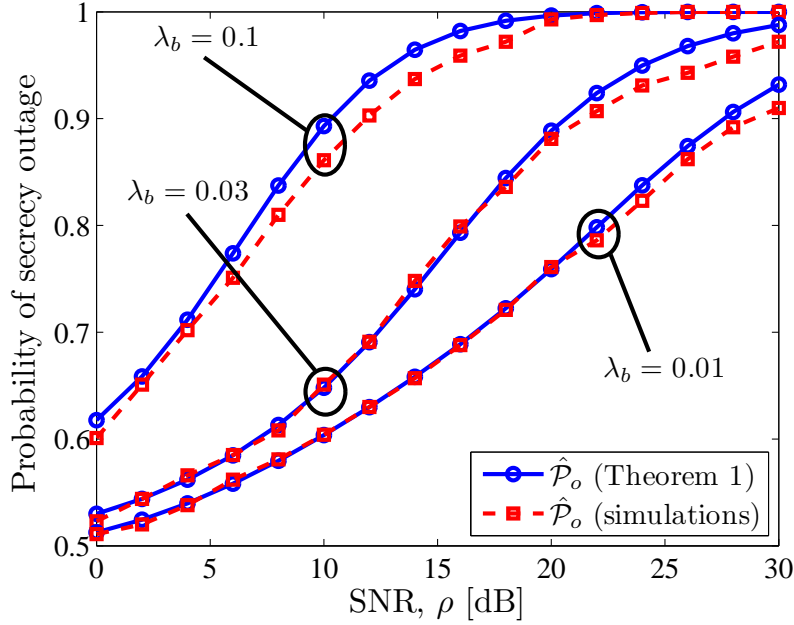}
\caption{Comparison between the simulated probability of secrecy outage $\hat{\mathcal{P}}_o$ and the analytical result from Theorem 1, for $N = 20$ transmit antennas, $K=20$ users per BS, and three values of the density of BSs $\lambda_b$.}
\label{fig:Outage_ana_sim}
\end{figure}

In Fig. \ref{fig:Outage_ana_sim} we compare the simulated probability of secrecy outage $\hat{\mathcal{P}}_o$ to the analytical result given in Theorem 1, for $N = 20$ transmit antennas, $K=20$ users per BS, and three values of the density of BSs $\lambda_b$. The analytical curves were obtained by using lognormal approximations for the pdfs $f_{\hat{I}}(x,y)$ and $f_{\hat{L}}(z)$. The figure shows that the result provided in Theorem 1 is accurate for all values of $\lambda_b$ at relatively low values of SNR. Due to the lognormal approximations, the result is slightly less accurate at relatively high values of SNR, when the BS can cancel the intra-cell interference and leakage \cite{GeraciJSAC}, and the secrecy outage is mostly determined by $\hat{I}$ and $\hat{L}$.

\begin{figure}
\centering
\includegraphics[width=\figwidth]{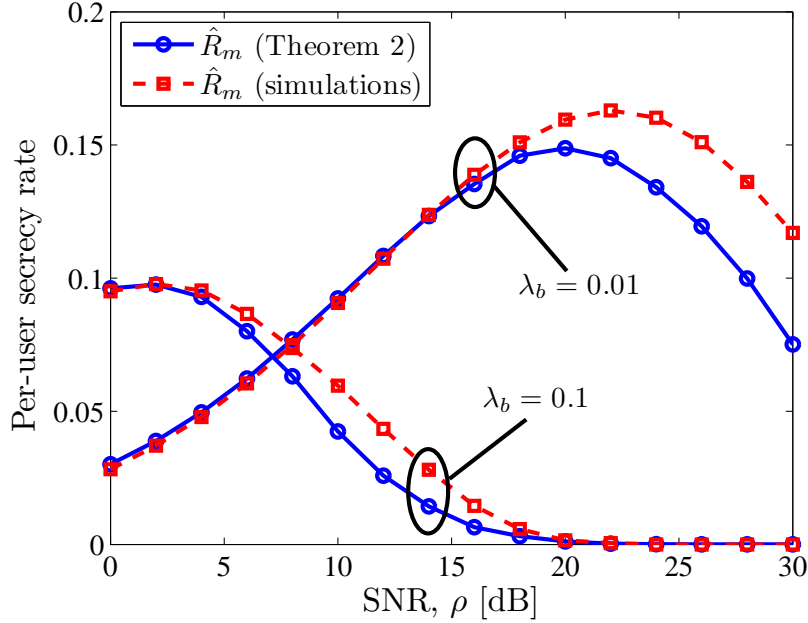}
\caption{Comparison between the simulated mean secrecy rate $\hat{R}_m$ and the analytical result from Theorem 2, for $N = 20$ transmit antennas, $K=20$ users per BS, and two values of the density of BSs $\lambda_b$.}
\label{fig:Rate_ana_sim}
\end{figure}

In Fig. \ref{fig:Rate_ana_sim} we compare the simulated mean secrecy rate $\hat{R}_m$ to the analytical result given in Theorem 2, for $N = 20$ transmit antennas, $K=20$ users per BS, and two values of the density of BSs $\lambda_b$. The analytical curves were again obtained by using lognormal approximations for the pdfs $f_{\hat{I}}(x,y)$ and $f_{\hat{L}}(z)$. The figure shows that the simulations and the analytical result from Theorem 2 follow the same trend. Therefore, the analysis provides insights into the behavior of the secrecy rate as a function of $\lambda_b$ and the SNR. The result from Theorem 2 is accurate for all values of $\lambda_b$ at relatively low values of SNR. Again due to the lognormal approximations, the analytical curve is less accurate at relatively high values of SNR, since $\chi$ vanishes as reported in (\ref{eqn:chi}), and the secrecy rate in (\ref{eqn:mean_rate_generic}) is dominated by $\hat{I}$ and $\hat{L}$. This inaccuracy could be avoided by employing the exact pdfs of $f_{\hat{I}}(x,y)$ and $f_{\hat{L}}(z)$, obtained from their Laplace transforms in Lemma \ref{Lemma:LaplaceI} and Lemma \ref{Lemma:LaplaceL}, but the anti-transform operation is computationally expensive. Finding better approximations for $f_{\hat{I}}(x,y)$ and $f_{\hat{L}}(z)$ is therefore identified as a promising research problem.

\begin{figure}
\centering
\includegraphics[width=\figwidth]{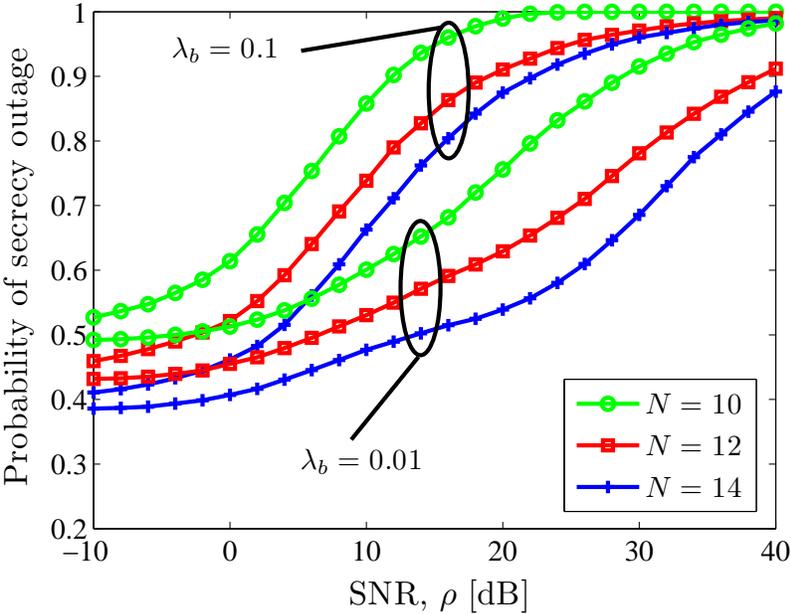}
\caption{Simulated probability of secrecy outage versus SNR, for $K=10$ users per BS and various values of the number of transmit antennas $N$ and density of BSs $\lambda_b$.}
\label{fig:Outage_vs_SNR_N}
\end{figure}

In Fig. \ref{fig:Outage_vs_SNR_N} we plot the simulated probability of secrecy outage versus the SNR, for $K=10$ users per BS and three values of the number of transmit antennas $N$. In this figure, two cases are considered for the density of BSs $\lambda_b$, namely $0.01$ and $0.1$, while the density of users is given by $\lambda_u = K \lambda_b$. Figure \ref{fig:Outage_vs_SNR_N} shows that RCI precoding achieves confidential communications in cellular networks with probability of secrecy outage $\hat{\mathcal{P}}_o < 1$, and that having more transmit antennas is beneficial as it reduces the probability of secrecy outage. However unlike an isolated cell \cite{GeraciTWC}, cellular networks tend to be in secrecy outage w.p. $1$ if the transmit power grows unbounded, irrespective of the number of transmit antennas. These observations are consistent with Remark \ref{R:outage}.

\begin{figure}
\centering
\includegraphics[width=\figwidth]{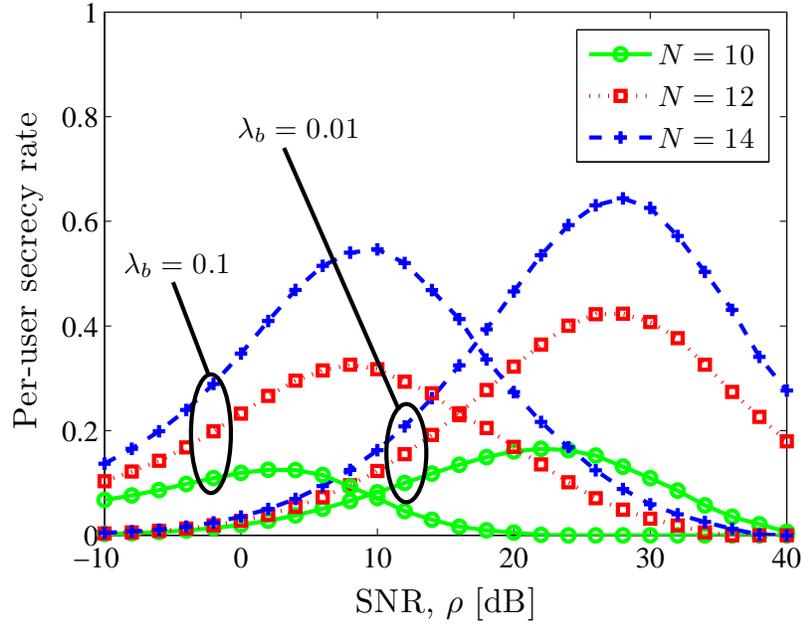}
\caption{Simulated ergodic secrecy rate versus SNR, for $K=10$ users per BS and various values of the number of transmit antennas $N$ and density of BSs $\lambda_b$.}
\label{fig:Rate_vs_SNR_N}
\end{figure}

In Fig. \ref{fig:Rate_vs_SNR_N} we plot the simulated per-user ergodic secrecy rate versus the SNR, for $K=10$ users per BS and three values of the number of transmit antennas $N$. In this figure, again, two cases are considered for the density of BSs $\lambda_b$, namely $0.01$ and $0.1$, while the density of users is given by $\lambda_u = K \lambda_b$. Fig. \ref{fig:Rate_vs_SNR_N} shows that in cellular networks RCI precoding can achieve a non-zero secrecy rate, and that having more transmit antennas is beneficial as it increases the secrecy rate. However unlike the secrecy rate in an isolated cell \cite{GeraciJSAC}, the secrecy rate in a cellular scenario does not grow unbounded with the SNR, even with a large number of transmit antennas. These observations are consistent with Remark \ref{R:SNR}.

\begin{figure}
\centering
\includegraphics[width=\figwidth]{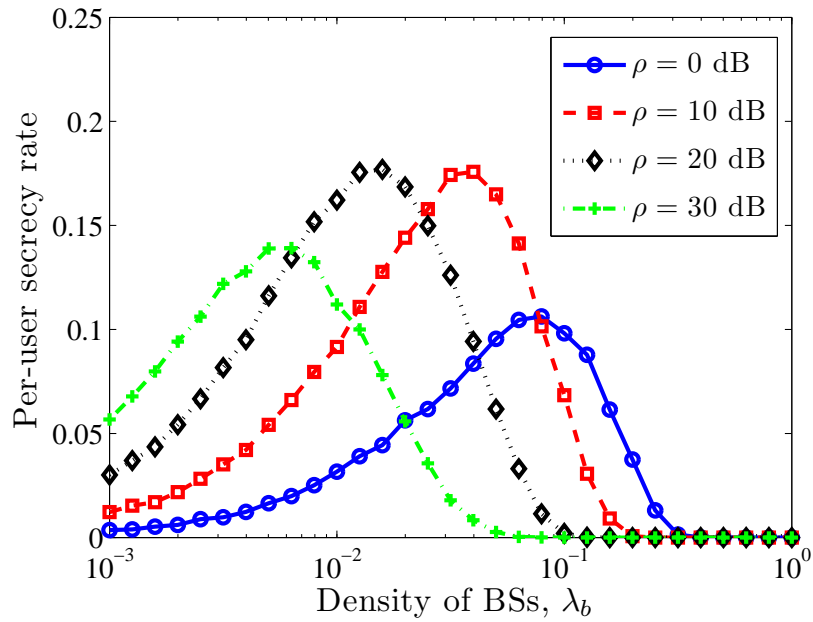}
\caption{Simulated ergodic secrecy rate versus density of BSs, for $N = 20$ transmit antennas, $K=20$ users per BS, and various values of the SNR $\rho$.}
\label{fig:Rate_vs_lambdab}
\end{figure}

In Fig. \ref{fig:Rate_vs_lambdab} we plot the simulated per-user ergodic secrecy rate as a function of the density of BSs $\lambda_b$, for $N = 20$ transmit antennas, $K=20$ users per BS, and various values of the SNR. Fig. \ref{fig:Rate_vs_lambdab} shows that there is an optimal value for the density of BSs $\lambda_b$ that maximizes the secrecy rate, and that such value is smaller for higher values of the SNR. This observation is consistent with Remark \ref{R:lambda}.

\section{Conclusion}
In this paper, we considered physical layer security for the downlink of cellular networks, where multiple base stations (BSs) generate inter-cell interference, and malicious users of neighboring cells can cooperate to eavesdrop. We showed that RCI precoding can achieve a non-zero secrecy rate with probability of outage smaller than one. However we also found that unlike isolated cells, the secrecy rate in a cellular network does not grow monotonically with the signal-to-noise ratio (SNR), and the network tends to be in secrecy outage if the transmit power grows unbounded. We further showed that there is an optimal value for the density of BSs that maximizes the secrecy rate, and this value is a decreasing function of the SNR. Using the developed analysis, we clearly established the importance of designing the transmit power and the BS deployment density to make communications robust against malicious users in other cells. 

Investigating the secrecy rates in heterogeneous networks, where small BSs are overlaid within the macro cellular network based on traffic and coverage demand, is identified as a promising research problem. Moreover, obtaining an exact characterization of the pdfs of the interference and leakage power could be an interesting future research direction.
\appendices

\section{}
\begin{IEEEproof}[Proof of Proposition \ref{Proposition:distributions}]
Under RCI precoding, the BS $b$ multiplies the confidential message $m_{b,j}$ destined for user $j$, for $1 \leq j \leq K_b$, by $\mathbf{w}_{b,j}$, so that the transmitted signal is a linear function of the message $m_{b,j}$, i.e., $\mathbf{x}_b = \sqrt{P} \sum_{j=1}^{K_b} \mathbf{w}_{b,j} m_{b,j}$. The inter-cell interference power gain at the typical user $o$ is given by $g_{b,o} = \sum_{j=1}^{K_b} \left|  \mathbf{h}_{b,o}^{\dagger} \tilde{\mathbf{w}}_{b,j} \right|^2$, with $\tilde{\mathbf{w}}_{b,j} = \sqrt{K_b} \mathbf{w}_{b,j}$. The normalized precoding vectors $\tilde{\mathbf{w}}_{b,j}$ have unit-norm on average, and they are calculated independently of $\mathbf{h}_{b,o}^{\dagger}$. Therefore, $\mathbf{h}_{b,o}^{\dagger}$ and $\tilde{\mathbf{w}}_{b,j}$ are independent isotropic unit-norm random vectors, and $\left|  \mathbf{h}_{b,o}^{\dagger} \tilde{\mathbf{w}}_{b,j} \right|^2$ is a linear combination of $N$ complex normal random variables, i.e., exponentially distributed. As a result, we have that $g_{b,o} \sim \Gamma(K_b,1)$, since it is the sum of $K_b$ i.i.d. exponential r.v. 

The leakage power gain at the malicious user $e \in \mathcal{M}_o^E$ is given by $g_{c,e} = \left|  \mathbf{h}_{c,e}^{\dagger} \tilde{\mathbf{w}}_{c,o} \right|^2$, with $\tilde{\mathbf{w}}_{c,o} = \sqrt{K_c} \mathbf{w}_{c,o}$. Similarly, we have that $\mathbf{h}_{c,e}^{\dagger}$ and $\tilde{\mathbf{w}}_{c,o}$ are independent isotropic unit-norm random vectors. As a result, we have that $g_{c,e} \sim \mathrm{exp}(1)$ since it is a linear combination of $N$ complex normal r.v.
\end{IEEEproof}

\section{}
\begin{IEEEproof}[Proof of Lemma \ref{Lemma:LaplaceI}]
The Laplace transform of the interference term $\hat{I}$ can be derived as follows
\begin{align}
\mathbb{E} \left[ e^{-s\hat{I}} \right] &= \mathbb{E} \left[ e^{- \frac{s}{K} \sum_{b \in \Phi_b \backslash c} g_{b,o} \|b\|^{-\eta} } \right] = \mathbb{E} \left[ \prod_{b \in \Phi_b \backslash c} e^{-\frac{s}{K} g_{b,o} \|b\|^{-\eta} } \right] \stackrel{(a)}{=} \mathbb{E}_{\Phi_b} \left[ \prod_{b \in \Phi_b \backslash c} \mathcal{L}_{g_{b,o}} \left( \frac{s}{K} \|b\|^{-\eta} \right) \right] \nonumber\\
&\stackrel{(b)}{=} \textrm{exp} \left\{ - \lambda_b \int_{\mathbb{R}^2 \cap \bar{\mathcal{B}}(o,\|c\|)} \left[ 1 - \mathcal{L}_{g_{b,o}} \left( \frac{s}{K} \|b\|^{-\eta} \right) \right] \textrm{d} b \right\} \nonumber\\
&\stackrel{(c)}{=} \textrm{exp} \left\{ - \lambda_b \int_{\mathbb{R}^2 \cap \bar{\mathcal{B}}(o,\|c\|)} \left[ 1 - \frac{1}{\left(1 + \frac{s}{K} \|b\|^{-\eta} \right)^{K}} \right] \textrm{d} b \right\} \nonumber\\
&= \textrm{exp} \left\{ - \lambda_b \int_{\mathbb{R}^2 \cap \bar{\mathcal{B}}(o,\|c\|)} \frac{\left(1 + \frac{s}{K} \|b\|^{-\eta} \right)^{K} - 1}{\left(1 + \frac{s}{K} \|b\|^{-\eta} \right)^{K}} \textrm{d} b \right\} \nonumber\\
&\stackrel{(d)}{=} \textrm{exp} \left\{ - \lambda_b \int_{\mathbb{R}^2 \cap \bar{\mathcal{B}}(o,\|c\|)} \frac{\sum_{n=1}^{K} \binom{K}{n} \left( \frac{s}{K} \|b\|^{-\eta} \right)^{n}}{\left(1 + \frac{s}{K_b} \|b\|^{-\eta} \right)^{K}} \textrm{d} b \right\} \nonumber\\
&= \textrm{exp} \left\{ - \lambda_b \sum_{n=1}^{K} \binom{K}{n} \int_{\mathbb{R}^2 \cap \bar{\mathcal{B}}(o,\|c\|)} \frac{ \left( \frac{s}{K} \|b\|^{-\eta} \right)^{n}}{\left(1 + \frac{s}{K} \|b\|^{-\eta} \right)^{K}} \textrm{d} b \right\} \nonumber\\
&\stackrel{(e)}{=} \textrm{exp} \left\{ -2 \pi \lambda_b \left(\frac{s}{K}\right)^{\frac{2}{\eta}} \sum_{n=1}^{K} \binom{K}{n} \int_{\|c\|(\frac{s}{K})^{-\frac{1}{\eta}}}^{\infty} \frac{\nu^{-n \eta}}{\left(1 + \nu^{-\eta} \right)^{K}} \nu \textrm{d} \nu \right\} \nonumber\\
&\stackrel{(f)}{=} \mathrm{exp} \left\{ - \lambda_b \left(\frac{s}{K}\right)^{\frac{2}{\eta}} C_{\eta,K}\left(s,\|c\|\right) \right\},
\end{align}
where $(a)$ follows since the channel powers $g_{b,o}$ are independent of the locations of the BSs, $(b)$ follows from the PGFL of a PPP \cite{Stoyan96}, $(c)$ follows from the Laplace transform of $g_{b,o} \sim \Gamma(K,1)$, $(d)$ follows from the Binomial theorem, $(e)$ follows by converting to polar coordinates, and $(f)$ follows by substituting $(1+\nu^{-\eta})^{-1} = t$ and noting that the integral is the difference of two incomplete Beta functions $B(x;y,z) = \int_0^x t^{y-1}(1-t)^{z-1} \textrm{d} t$.
\end{IEEEproof}

\section{}
\begin{IEEEproof}[Proof of Proposition \ref{Proposition:lognormal}]
The mean and variance of the interference can be obtained by applying Campbell's theorem and are given by \cite{BaccelliBook1}
\begin{align}
\mu_{\hat{I}} &= \mathbb{E} \left[ \hat{I} \right] \stackrel{(a)}{=} \mathbb{E}_{\Phi_b} \left[ \sum_{b \in \Phi_b \backslash c} \|b\|^{-\eta} \right] = 2 \pi \lambda_b \int_{\|c\|}^{\infty} {v^{-\eta} v \,\mathrm{d}v } = \frac{2 \pi \lambda_b \|c\|^{-(\eta-2)}}{\eta-2} \\
\sigma^2_{\hat{I}} &= \mathbb{E} \left[ \hat{I}^2 \right] - \mu_{\hat{I}}^2 \stackrel{(b)}{=} \frac{K+K^2}{K^2} \mathbb{E} \left[ \sum_{b \in \Phi_b \backslash c} \|b\|^{-2\eta} \right] \nonumber\\
&= \frac{2 \pi \lambda_b \left(K+K^2\right)}{K^2} \int_{\|c\|}^{\infty} v^{-2\eta} v \, \mathrm{d}v = \frac{\pi \lambda_b \left(K+K^2\right) \|c\|^{-2(\eta-1)}}{K^2 \left(\eta-1\right)},
\end{align}
where $(a)$ follows from $\mathbb{E}[g_{b,o}]=K$, and $(b)$ follows from $\mathbb{E}\left[g^2_{b,o}\right]=K+K^2$.
Similarly, the mean and variance of the leakage are given by \cite{BaccelliBook1}
\begin{align}
\mu_{\hat{L}} &= \mathbb{E} \left[ \hat{L} \right] \stackrel{(c)}{=} \frac{1}{K} \mathbb{E}_{\Phi_u} \left[ \sum_{e \in \hat{\mathcal{M}}_o^E } \mathbb{E} \left[ g_{c,e} \right] \|e-c\|^{-\eta} \right] = \frac{2 \pi \lambda_u}{K} \int_r^{\infty} {v^{-\eta} v \,\mathrm{d}v } = \frac{2 \pi \lambda_u r^{-(\eta-2)}}{K \left(\eta-2\right)} \\
\sigma^2_{\hat{L}} &= \mathbb{E} \left[ \hat{L}^2 \right] - \mu_{\hat{L}}^2 \stackrel{(a)}{=} \frac{2}{K^2} \mathbb{E} \left[ \sum_{e \in \hat{\mathcal{M}}_o^E } \|e-c\|^{-2\eta} \right] =  \frac{4 \pi \lambda_u }{K^2} \int_r^{\infty} v^{-2\eta} v \, \mathrm{d}v = \frac{2 \pi \lambda_u r^{-2(\eta-1)}}{K^2 \left(\eta-1\right)},
\end{align}
where $(c)$ follows from $\mathbb{E}[g_{c,e}]=1$, and $(d)$ follows from $\mathbb{E}\left[g^2_{c,e}\right]=2$.
\end{IEEEproof}

\section{}
\begin{IEEEproof}[Proof of Theorem \ref{T:mean_rate}]
By using approximation (\ref{eqn:R_hat}) in (\ref{eqn:mean_rate_definition}), we obtain
\begin{align}
R_m &\approx \mathbb{E} \left[ \hat{R} \right] = \mathbb{E} \Bigg[ \Bigg\{ \log_2 \Bigg( 1 + \frac {\rho \alpha \|c\|^{-\eta}} {\rho \chi \|c\|^{-\eta} + \rho \hat{I} + 1 } \Bigg) - \log_2 \Bigg( 1 + \rho \chi \|c\|^{-\eta} + \rho \hat{L} \Bigg) \Bigg\}^+ \Bigg] \nonumber \\
&= \mathbb{E} \Bigg[ \Bigg[ \log_2 \Bigg( 1 + \frac {\rho \alpha \|c\|^{-\eta}} {\rho \chi \|c\|^{-\eta} + \rho \hat{I} + 1 } \Bigg) - \log_2 \Bigg( 1 + \rho \chi \|c\|^{-\eta} + \rho \hat{L} \Bigg) \Bigg] \indic_{(\hat{L}<\tau(\hat{I},\|c\|))} \Bigg] \nonumber \\
&= \underset{\hat{I},\|c\|}{\mathbb{E}} \Bigg[ \log_2 \Bigg( 1 + \frac {\rho \alpha \|c\|^{-\eta}} {\rho \chi \|c\|^{-\eta} + \rho \hat{I} + 1 } \Bigg) \mathbb{P} \left( \hat{L} < \tau \left( \hat{I},\|c\| \right) \right) \Bigg] \nonumber\\
& \quad - \underset{\hat{I},\hat{L},\|c\|}{\mathbb{E}} \Bigg[ \log_2 \Bigg( 1 + \rho \chi \|c\|^{-\eta} + \rho \hat{L} \Bigg) \indic_{(\hat{L}<\tau(\hat{I},\|c\|))} \Bigg] \nonumber \\
& \stackrel{(a)}{=} \int_0^{\infty} \int_{-\infty}^{\frac{\alpha}{\rho \chi}-\frac{1}{\rho} - \chi y^{-\eta}} \left\{ \log_2 \Bigg( 1 + \frac {\rho \alpha y^{-\eta}} {\rho \chi y^{-\eta} + \rho x + 1 } \Bigg) \int_{-\infty}^{\tau (x,y)} f_{\hat{L}}(z) \right. \nonumber \\
& \quad \left. - \int_{-\infty}^{\tau (x,y)} \log_2 \left( 1 + \rho \chi y^{-\eta} + \rho z \right) \, f_{\hat{L}}(z) \, \mathrm{d} z \right\} \, f_{\hat{I}}(x,y) \, \mathrm{d} x \, 2 \lambda_b \pi y e^{-\lambda_b \pi y^2} \, \mathrm{d} y,
\end{align}
where $\indic_{(\cdot)}$ is the indicator function, and where the upper limit in the inner integration in $(a)$ follows from $0 \leq \hat{L} < \tau (\hat{I},\|c\|)$.
\end{IEEEproof}

\section{}
\begin{IEEEproof}[Proof of Corollary \ref{C:Plancherel}]
The lower bound in (\ref{eqn:mean_rate_LB}) can be obtained as follows
\begin{align}
\hat{R}_m &= \mathbb{E} \left[ \hat{R} \right] \stackrel{(a)}{\geq} \Bigg\{ \mathbb{E} \Bigg[ \log_2 \Bigg( 1 + \frac {\rho \alpha \|c\|^{-\eta}} {\rho \chi \|c\|^{-\eta} + \rho \hat{I} + 1 } \Bigg) - \log_2 \Bigg( 1 + \rho \chi \|c\|^{-\eta} + \rho \hat{L} \Bigg) \Bigg] \Bigg\}^+ \nonumber\\
&= \Bigg\{ \int_0^{\infty} \int_{-\infty}^{\infty} \Bigg[ \log_2 \Bigg( 1 + \frac {\rho \alpha y^{-\eta}} {\rho \chi y^{-\eta} + \rho x + 1 } \Bigg) f_{\hat{I}}(x,y) \nonumber\\
& \quad - \log_2 \Bigg( 1 + \rho \chi y^{-\eta} + \rho x \Bigg) f_{\hat{L}}(x) \Bigg] \mathrm{d} x \, 2 \lambda_b \pi y e^{- \lambda_b \pi y^2} \mathrm{d} y \Bigg\}^+ \nonumber \\ 
&\stackrel{(b)}{=} \Bigg\{ \int_0^{\infty} \int_{-\infty}^{\infty} \Bigg[ \mathcal{F}^*_1(\phi,y) \mathcal{L}_{\hat{I}}(-i 2 \pi \phi,y) - \mathcal{F}^*_2(\phi,y) \mathcal{L}_{\hat{L}}(-i 2 \pi \phi) \Bigg] \mathrm{d} \phi \, 2 \lambda_b \pi y e^{- \lambda_b \pi y^2} \mathrm{d} y \Bigg\}^+.
\label{eqn:double_int}
\end{align}
Equation $(a)$ follows from Jensen's inequality $\mathbb{E}[x^+] \geq \{\mathbb{E}[x]\}^+$. Equation $(b)$ follows by Parseval's theorem \cite{Rudin}, and since
\begin{align}
\mathcal{F}_1(\phi,y) &= \frac{\mathrm{sgn}(\phi) \, e^{2 \pi i (\chi y^{-\eta} + \frac{1}{\rho}) \phi}}{2 \phi \log 2}\left(1 - e^{2 \pi i \alpha \phi y^{-\eta}} \right)\\
\mathcal{F}_2(\phi,y) &= \frac{-e^{2 \pi i (\chi y^{-\eta} + \frac{1}{\rho}) \phi}}{\log 2} \left[ \frac{1}{2\left|\phi\right|} + \frac{\gamma}{\rho} \delta(\phi) \right]
\end{align}
are the respective Fourier transforms of 
\begin{align}
f_1(x,y) &= \log_2 \Bigg( 1 + \frac {\rho \alpha y^{-\eta}} {\rho \chi y^{-\eta} + \rho x + 1 } \Bigg) \\
f_2(x,y) &= \log_2 \Bigg( 1 + \rho \chi y^{-\eta} + \rho x \Bigg),
\end{align}
where $\gamma = \lim_{n \rightarrow \infty} \left(\sum_{k=1}^{n} \frac{1}{k} - \log n \right)$ is the Euler-Mascheroni constant. The functions $\mathcal{F}_1(\phi,y)$ and $\mathcal{F}_2(\phi,y)$ can be obtained from the Fourier transforms of $\frac{1}{x}$ and $\log|x|$, and by applying the differentiation and shift theorems. 
\end{IEEEproof}

\ifCLASSOPTIONcaptionsoff
  \newpage
\fi
\bibliographystyle{IEEEtran}
\bibliography{Bib_Giovanni}
\end{document}